\documentclass{ws-ijmpe}
\usepackage[super,compress]{cite}
\usepackage{lineno,hyperref}
\begin{document}

\markboth{Kanhaiya Jha \textit{et al.}}{Modification of tensor force in \textit{p}-shell effective interaction}

\catchline{}{}{}{}{}
\title{Modification of tensor force in \textit{p}-shell effective interaction}

\author{Kanhaiya Jha\footnote{E-mail: kanhaiya.jha@iitrpr.ac.in},  Pawan Kumar, Shahariar Sarkar and P. K. Raina}

\address{Department of Physics,\\ 
	     Indian Institute of Technology Ropar,\\ 
	     Rupnagar, Punjab,  140001, India }

\maketitle

\begin{history}
\received{Day Month Year}
\revised{Day Month Year}
\end{history}

\begin{abstract}
In many shell model interactions, the tensor force monopole matrix elements often retain systematic trends originating in the bare tensor force. However,  in the present work, we find that Isospin T = 0 tensor force monopole matrix elements of \textit{p}-shell effective interaction  CK(8-16) do not share these systematic. We correct these discrepancies by modifying T = 0 tensor force two-body matrix elements (TBMEs) of CK(8-16) by the analytically calculated tensor force TBMEs. With some additional modification of single-particle energies and TBMEs, the revised effective interaction is named as CKN. The effective interaction CKN has been tested for the calculations of \textit{p}-shell nuclei of normal parity states from various physics viewpoints such as excitation spectra, electromagnetic moments, and electromagnetic and \textit{G}amow-\textit{T}eller (\textit{GT}) transitions. The obtained results are found to be satisfactory.
	
\end{abstract}

\keywords{Shell model; effective interaction; spin-tensor decomposition; tensor force.}

\ccode{PACS numbers:}


\section{Introduction}
\label{sec1}

The success of the nuclear shell model mainly depends on the employed effective two-nucleon interaction which is generally derived using bare  NN interaction \cite{brown, brown1, caurier, otsuka1, jensen}. However, the effective interaction made this way is further adjusted empirically to incorporate the in-medium effect, and many-body force effect to successfully explain the nuclear structural properties \cite{jensen, kuo, dean, chung}. For example, the interactions USD \cite{usd} and GXPF1 \cite{honma} are derived jointly by ETBME+\textit{G}
hamiltonians \cite{brown1}. The monopole \cite{zuker, dufour} modification of the effective interaction is one of the most extensively used empirical methods which significantly improve the description of experimental data  \cite{poves, poves2, otsuka2, honma1, honma2}.

In recent years, the importance of individual components of NN effective interaction i.e., central, spin-orbit and tensor forces to the single-particle energy gaps and shell evolution in neutron-rich nuclei have been extensively explored \cite{otsuka2, umeya1, otsuka3,umeya,otsuka4,smirnova,otsuka5,pawan}. The tensor force of bare NN interaction in particular gains lot of interest due to its unique and robust features. It plays an important role in the evolution of shell structure throughout the nuclear landscape as demonstrated by Otsuka and his collaborators \cite{otsuka3, otsuka6}. The tensor force monopole matrix element $\bar{V}^{T}_{jj'} (\zeta)$ is attractive for the configuration $j_{>} j'_{<}$\footnote{$j_{>}= l+\frac{1}{2}$, $j_{<}= l-\frac{1}{2}$} ($j_{<} j'_{>}$) and repulsive for  the configuration $j_{>} j'_{>}$ ($j_{<} j'_{<}$) \cite{otsuka3, umeya}. Furthermore, the tensor force monopole matrix elements are barely change against the various renormalization procedures, and persists their unique systematic properties \cite{tsunoda}. The numerical study based on the spin-tensor decomposition (STD) \cite{kirson, klingenback, kenji} also reveal that the $\bar{V}^{T}_{jj'} (\zeta)$ matrix elements of well established effective shell-model interaction USDB \cite{brown2} has the same property as of the bare tensor force \cite{wang, wang1}. In the present work, we are doing the similar analysis to investigate the $\bar{V}^{T}_{jj'} (\zeta)$ matrix elements of widely used \textit{p}-shell effective interaction CK(8-16) \cite{cohen}. In our investigation,  we find that the systematics properties of tensor force is present for Isospin T = 1 in CK(8-16), however, the same properties are missing for their T = 0 tensor force matrix elements. The tensor force matrix elements $\bar{V}_{p3p3}^{T = 0}$, $\bar{V}_{p3p1}^{T = 0}$, and $\bar{V}_{p1p1}^{T = 0}$ of the interaction is found to be opposite than the expected. The other \textit{p}-shell effective interactions CKPOT \cite{cohen} and PTBME \cite{julies} are also lacking to reproduce the systematic features of tensor force for T = 0. A similar kind of discrepancies have been also reported recently for T =  1 tensor force matrix elements of \textit{pf}-shell effective interactions of GX-family \cite{wang, kjha}.

The effective interaction CK(8-16) discussed above was constructed phenomenologically by optimizing TBMEs and single-particle energies to successfully explain the experimental observables. As a  consequence, the peculiar character of the T = 0 tensor force may originate in the
CK(8-16). The tensor force of CK(8-16) is inappropriate, hence, there is a possibility that other forces, i.e., central and spin-orbit forces, are inappropriate too. Since the systematic properties of spin-orbit force are not yet known, and \textit{pp}-matrix elements of central force are following the same properties as reported in the literature \cite{kjha}, we do not change these forces at the moment. The comprehensive study by changing these forces will be fascinating in this field. In the present work, we pay attention to the tensor force discrepancies present in the CK(8-16). The modification is done using the semi-empirical method, discussed later in the article in detail. To assess the credibility of the revised interaction, we have done calculations for \textit{p}-shell nuclei of normal parity i.e., $\pi = (-1)^{A}$ states with various physics viewpoints such as excitation spectra, electromagnetic moments, and electromagnetic and \textit{G}amow-\textit{T}eller transitions.

This paper is organized as follows. In Sec.\ref{sec2}, we have
discussed the theoretical framework to obtain the monopole matrix elements of the interactions, and detail formalism to correct the tensor force discrepancies present in the effective interaction. Results and discussion are presented in Sec.\ref{sec3}, and finally, a summary of the work is given in Sec.~\ref{sec4}.

\section{Theoretical framework}
\label{sec2}
\subsection{Spin-tensor decomposition}

The spin-tensor decomposition (STD) \cite{kirson, klingenback, kenji} is a useful tool to separate the NN effective interaction into its central, spin-orbit and tensor force structure. The STD has been used in various recent studies \cite{umeya, smirnova, pawan, sarkar} to explore the role of individual components of NN interactions in the shell evolution of neutron-rich nuclei. In the present study, we have used the STD tool to find the discrepancies present in the \textit{p}-shell effective interaction CK(8-16). 

In spin-tensor decomposition, the interaction between two-nucleon is defined as the linear sum of the scalar product of configuration space operator \textit{Q} and spin space operator \textit{S} of rank k:
\begin{equation}
	V= \sum_{k = 0}^{2} V(k) = \sum_{k = 0}^{2} Q^{k} .  S^{k} , 
\end{equation}
where, rank k = 0, 1, and 2 represent central, spin-orbit, and
tensor force, respectively. Since the TBME's of NN interaction are defined in \textit{jj}-basis, therefore, one has to transform it first from \textit{jj}-basis to \textit{LS}-coupling in a standard way using \textit{9j}-symbol relation, and then decomposed the effective interaction to the individual components based on their tensor structure. The final expression of V(k) obtained from the \textit{jj}-coupled matrix elements can be written as\\
\[
\begin{aligned}
<ab: JT|V(k)|cd: JT> ~ = ~ \frac{1}{\sqrt{(1+\delta_{n_{a}n_{b} l_{a} l_{b} j_{a} j_{b}}) (1+\delta_{n_{c}n_{d} l_{c} l_{d} j_{c} j_{d}})}} \\
\sum_{LSL'S'} \
\sqrt{(2j_{a}+1) (2j_{b}+1)(2L+1)(2S+1)}
\begin{Bmatrix}
l_{a} & 1/2 & j_{a}\\
l_{b} & 1/2 & j_{b}\\
L & S & J
\end{Bmatrix}\\
\sqrt{(2j_{c}+1) (2j_{d}+1)(2L'+1)(2S'+1)}
\begin{Bmatrix}
l_{c} & 1/2 & j_{c}\\
l_{d} & 1/2 & j_{d}\\
L' & S' & J
\end{Bmatrix}\\
(2k+1) (-1)^{J}
\begin{Bmatrix}
L & S & J\\
S' & L' & k
\end{Bmatrix}
\sum_{J'}(-1)^{J'} (2J'+1)
\begin{Bmatrix}
L & S & J'\\
S' & L' & k
\end{Bmatrix}\\
\sum_{j'_{a} j'_{b}j'_{c} j'_{d}}
\sqrt{(2j'_{a}+1) (2j'_{b}+1)(2L+1)(2S+1)}
\begin{Bmatrix}
l_{a} & 1/2 & j'_{a}\\
l_{b} & 1/2 & j'_{b}\\
L & S & J'
\end{Bmatrix}\\  
\sqrt{(2j'_{c}+1) (2j'_{d}+1)(2L'+1)(2S'+1)}
\begin{Bmatrix}
l_{c} & 1/2 & j'_{c}\\
l_{d} & 1/2 & j'_{d}\\
L' & S' & J'
\end{Bmatrix}\\  
\sqrt{(1+\delta_{n_{a}n_{b} l_{a} l_{b} j'_{a} j'_{b}}) (1+\delta_{n_{c}n_{d} l_{c} l_{d} j'_{c} j'_{d}})}
<\alpha \beta: JT|V|\gamma \delta : JT>\\
\end{aligned}
\]
where $a$ = $(n_{a} l_{a} j_{a})$ , $\alpha$ = $(n_{a} l_{a} j'_{a})$ are shorthand notation for the set of quantum numbers.

\subsection{Modification of effective interaction}

\begin{figure*}[ht!]
	\centering
	\includegraphics[height=10.0cm, width =14cm, trim={1.5cm 1.5cm -1.0cm 0.0cm}]{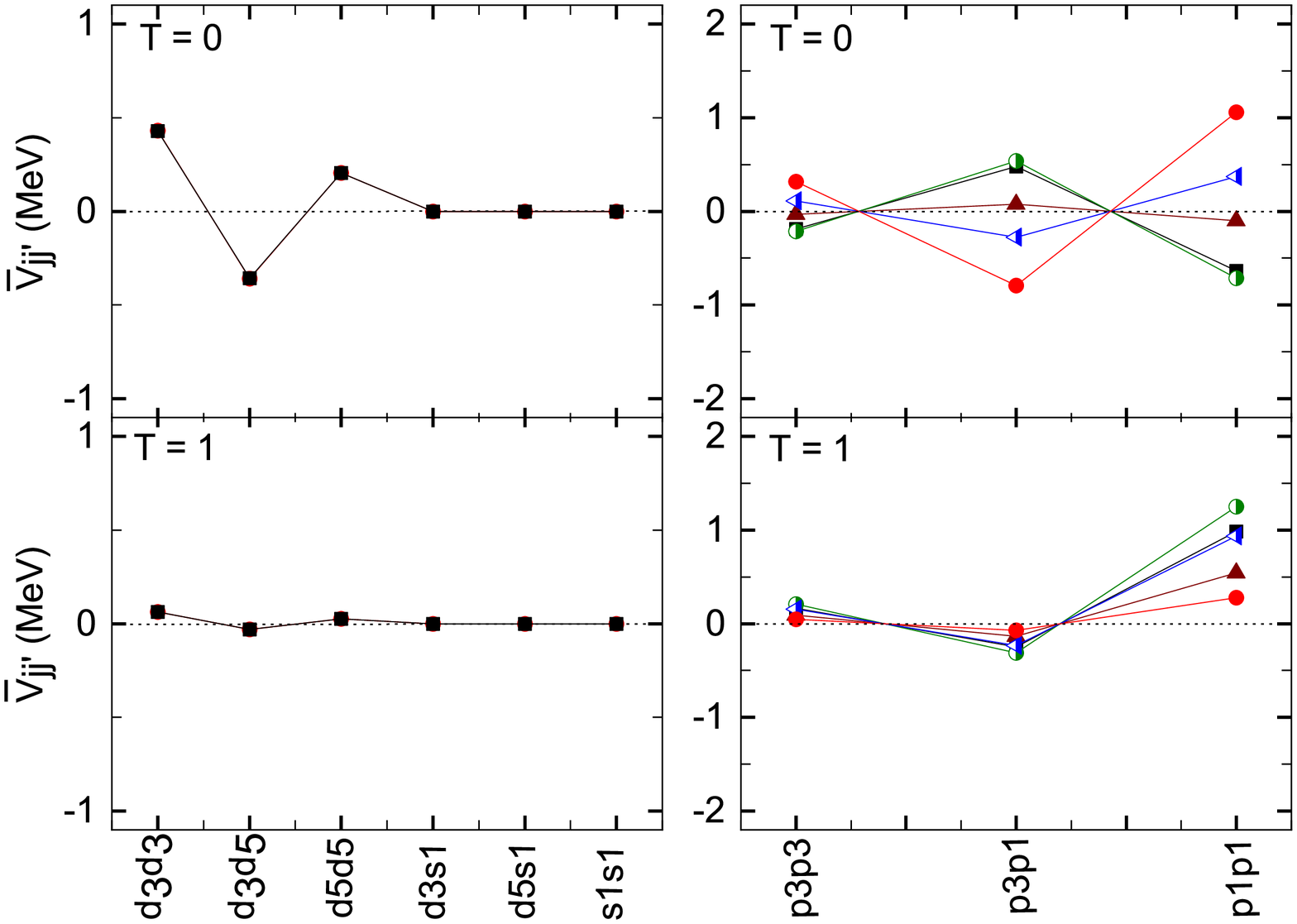}
	\caption{(Color online) Left side: Calculated (solid circle) tensor force matrix elements along with the USDB (solid square) interactions. Right side: Calculated (solid circle) tensor force monopole matrix elements along with the interactions CK(8-16) (solid square), CKPOT (half circle), PTBME (solid up triangle), and CKN (half triangle). Lines are drawn to guide the eyes.} 
	\label{F1}
\end{figure*}

The $\bar{V}^{T}_{jj'} (\zeta)$ matrix elements of shell model effective interaction CK(8-16) are obtained using spin-tensor decomposition for both Isospin T =  0 and 1, shown on the right side of Fig.~\ref{F1}. In T = 1 channel, all $\bar{V}^{T = 1}_{jj'} (\zeta)$ matrix elements are found to have their systematics properties, however, in T = 0 channel, we find irregularities. The matrix elements $\bar{V}_{p3p3}^{T = 0}$ and $\bar{V}_{p1p1}^{T = 0}$ are found attractive while $\bar{V}_{p3p1}^{T = 0}$ is found repulsive. However, it should be expected to be opposite i.e., $\bar{V}_{p3p3}^{T = 0}$ and $\bar{V}_{p1p1}^{T = 0}$ should be repulsive and $\bar{V}_{p3p1}^{T = 0}$ should be attractive. We have shown the T = 0 tensor force monopole matrix elements of interactions CKPOT and PTBME in the same figure, and found similar behavior in their T = 0 tensor force monopole matrix elements.

\begin{figure*}[ht!]
	\centering
	\includegraphics[height= 10.0cm, width =14cm, trim={1.5cm 1.5cm -1.0cm 0.0cm}]{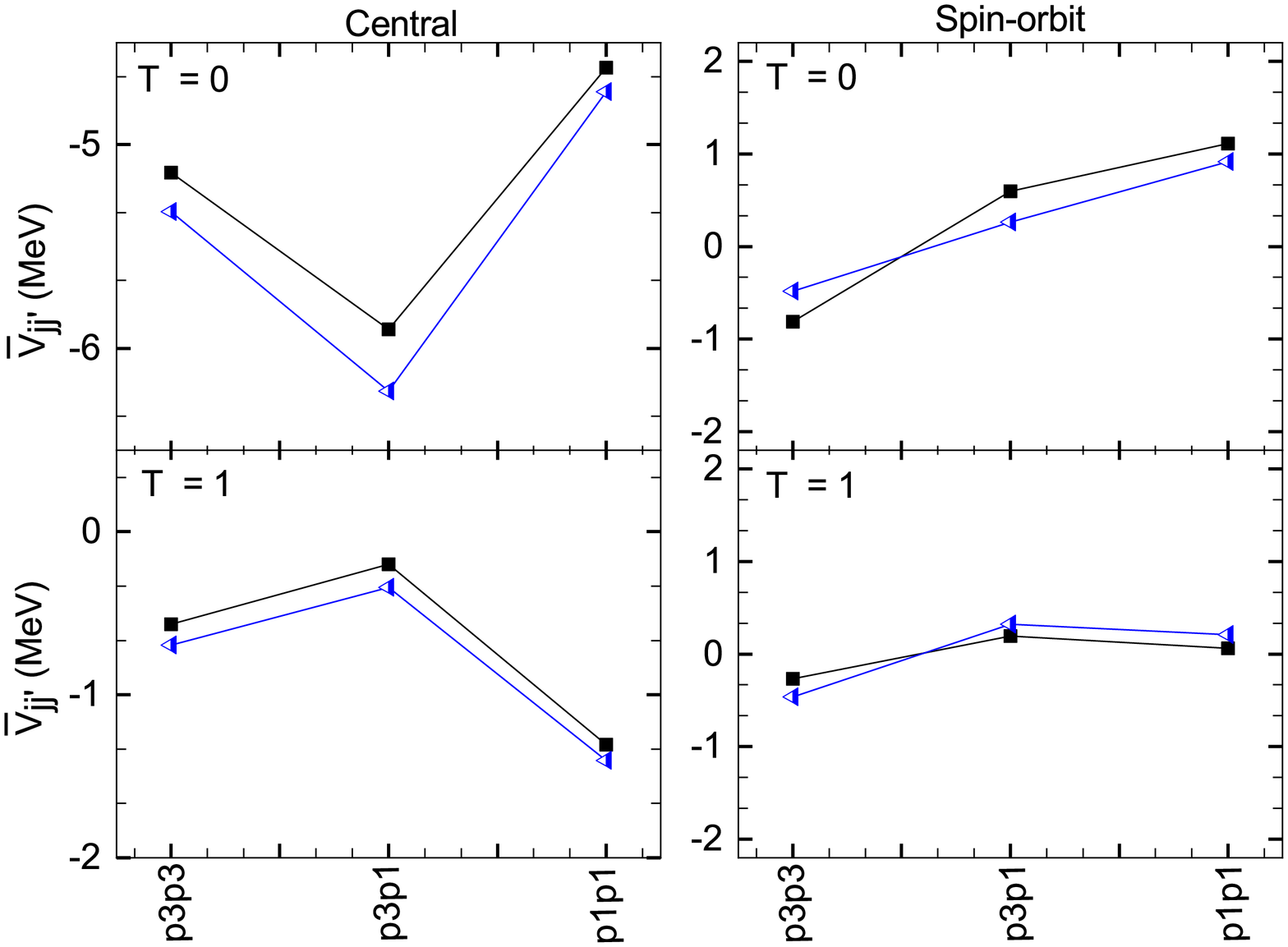}
	\caption{(Color online) Left side: Central force monopole matrix elements of interactions CKN (half triangle) and CK(8-16) (solid square) for both Isospin T = 0 and 1. Right side: Spin-orbit force monopole matrix elements of interactions CKN (half triangle) and CK(8-16) (solid square) for both Isospin T = 0 and 1. Lines are drawn to guide the eyes.} 
	\label{F2}
\end{figure*}

\begin{table}[ht!]
	\caption{\label{tab:table1}The individual components of TBME's of interaction CKN. Here $1 = 1p_{\frac{1}{2}}$ and $3 = 1p_{\frac{3}{2}}$. The unperturbed single-particle energy of $ 1p_{\frac{3}{2}}$ and $1p_{\frac{1}{2}}$ orbitals are 0.738 MeV and 2.23 MeV, respectively.}
	\label{tab1}
	\centering    
	\begin{tabular}{lcccccccccc}
		\cline{1-10}
		$j_{1}$ & $j_{2}$ & $j_{3}$ &$j_{4}$ & J & T & Total   & Central & Spin-orbit & Tensor \\
		\cline{1-10}
		3    &    3    &    3    &    3    &    1    &    0    &    -3.539    &    -3.689    &    0.051    &    0.100    \\
		3    &    3    &    3    &    3    &    3    &    0    &    -6.626    &    -6.031    &    -0.711    &    0.116    \\
		3    &    3    &    3    &    1    &    1    &    0    &    3.816    &    3.718    &    0.040    &    0.058    \\
		3    &    3    &    1    &    1    &    1    &    0    &    2.223    &    2.599    &    -0.160    &    -0.216    \\
		3    &    1    &    3    &    1    &    1    &    0    &    -6.472    &    -6.509    &    0.101    &    -0.064    \\
		3    &    1    &    3    &    1    &    2    &    0    &    -6.081    &    -6.031    &    0.355    &    -0.405    \\
		3    &    1    &    1    &    1    &    1    &    0    &    0.528    &    0.937    &    -0.381    &    -0.028    \\
		1    &    1    &    1    &    1    &    1    &    0    &    -3.457    &    -4.741    &    0.914    &    0.370    \\
		3    &    3    &    3    &    3    &    0    &    1    &    -3.193    &    -4.103    &    0.443    &    0.467    \\
		3    &    3    &    3    &    3    &    2    &    1    &    -0.567    &    -0.014    &    -0.646    &    0.093    \\
		3    &    3    &    3    &    1    &    2    &    1    &    -1.915    &    -1.851    &    0.002    &    -0.066    \\
		3    &    3    &    1    &    1    &    0    &    1    &    -4.865    &    -3.817    &    -0.387    &    -0.660    \\
		3    &    1    &    3    &    1    &    1    &    1    &    0.921    &    1.295    &    0.326    &    -0.700    \\
		3    &    1    &    3    &    1    &    2    &    1    &    -0.956    &    -1.323    &    0.320    &    0.047    \\
		1    &    1    &    1    &    1    &    0    &    1    &    -0.261    &    -1.404    &    0.209    &    0.934    \\
		\hline            
	\end{tabular}
\end{table}

In order to correct the discrepancies observed in CK(8-16) interaction, we have separately calculated T = 0 tensor force matrix elements using tensor force 
\begin{equation}
	V_{\zeta}= V(r)\sqrt{\frac{24 \pi}{5}}[Y^{(2)}.{(\sigma_{1} X  \sigma_{2} )}^{(2)}](\tau_{1} . \tau_{2}), 
\end{equation}
and replaced them with the tensor force matrix elements of CK(8-16). The radial dependency in the above expression of tensor force is treated with the Yukawa potential
\begin{equation}
	V(r) = -V_{0} \frac{e^{-r/a}}{r/a}, 
\end{equation} 
where $V_{0}$ is the strength parameter and ``\textit{a}'' is
the Compton scattering length of pion given as 1.41 fm for $m_{\pi}$ = 139.4 MeV. 

The $V_{0}$, in our calculation, is obtained from the fit of $\bar{V}^{T=0}_{jj'} (\zeta)$ matrix elements of USDB. On the left side of Fig.~\ref{F1}, the $\bar{V}^{T}_{jj'} (\zeta)$ calculated  analytically and USDB are shown for both T = 0 and 1. The analytically calculated tensor force matrix elements completely reproduces the tensor force matrix elements of USDB. The obtained $V_{0}$ for T = 0 is used to calculate $\bar{V}^{T= 0}_{jj'} (\zeta)$ matrix elements in $p$-shell. This comparison is shown on the right side of Fig.~\ref{F1}. The systematic properties of tensor force are found for all calculated tensor force matrix elements $\bar{V}_{p3p3}^{T = 0}$, $\bar{V}_{p3p1}^{T = 0}$ and $\bar{V}_{p1p1}^{T = 0}$. This approach is employed
recently to obtain the tensor force matrix elements in \textit{pf}-shell, and the obtained results are found to be in good agreement with the tensor force matrix elements of GX-family interactions \cite{kjha}.

The single-particle energies of $ 1p_{\frac{3}{2}}$ and $1p_{\frac{1}{2}}$ orbitals in CK(8-16) are 1.428 MeV and 1.570 MeV, respectively. These single-particle energies value are too low with respect to the \textit{G}-Matrix interaction \cite{jensen}. Many attempts had been made in past to enhance the gap between the $ 1p_{\frac{3}{2}}$ and $1p_{\frac{1}{2}}$ orbitals based on the observed spectra of $^{5}$He \cite{otsuka6, suzuki, suzuki1}. In the present case, we have set the single-particle energy $ \epsilon_{\frac{3}{2}}$ as 0.736 MeV which is the experimental one neutron separation energy of $^{5}$He with respect to the ground state $^{4}$He \cite{nndc, myo}. Further, the excitation energy of the $\frac{1}{2}_{1}^{-}$ state of $^{5}$He has very broad resonance with large error bar $\pm$ 1 MeV \cite{stevenson, coraggio} and it is taken around 1.50 MeV in the recent study \cite{myo}. Hence, we set the single-particle energy of $ \epsilon_{\frac{1}{2}}$ at 2.23 MeV to better reproduce the excitation energy of $^{5}$He.

After replacement of the T = 0 tensor force matrix elements and single-particle energies $ \epsilon_{\frac{3}{2}}$ and $ \epsilon_{\frac{1}{2}}$ of interaction CK(8-16), we have performed the shell model calculations. We find that the interaction after above changes nicely predicts spin of the ground and excited states, however, we find deviations in the predicted excitation energies in some of the cases. The 0$_{1}^{+}$ state of $^{10,12}$B, 2$_{1}^{+}$ state of $^{12}$B, $\frac{3}{2}_{1}^{-} $ state of $^{13}$C and 0$_{1}^{+}$ state of $^{14}$N are underpredicted by $0.78 - 2.16$ MeV with respect to the experimentally measured excitation spectra. In order to resolve these issues, we have done some additional modification. Since the unperturbed single-particle energy gap $ \epsilon_{\frac{1}{2}} - \epsilon_{\frac{3}{2}}$ has been increased by 1.35 MeV from the original CK(8-16) interaction, therefore, we have made $\bar{V}_{p3p1}^{T= 0}$ and $\bar{V}_{p1p1}^{T= 0}$ more attractive by the amount -0.125 MeV and -1.0 MeV, respectively. Moreover, there are some states discussed later in the article, which inappropriately predicted by interaction CK(8-16), have been improved by the modification of two-body matrix elements $V(2j_{a} 2j_{b} 2j_{c} 2j_{d}; JT) $: $V(3311; 10)$ and  $V(3333; 21) $ by 1.0 MeV and -0.4 MeV, respectively. It is interesting to note that the above modifications have improved the overall level structure discussed in the present work. Further, it is worth to mention that the two-body matrix elements modified in such a way that it does not affect the systematics properties of the tensor force. The derived interaction obtains this way, hereafter, is denoted by CKN. The individual force TBMEs of interaction CKN are shown in Table.~\ref{tab1}. The central and spin-orbit monopole components of CKN are shown in Fig.~\ref{F2} along with corresponding components of interaction CK(8-16). For T = 1, both central and spin-orbit components of CKN are almost equivalent to the original interaction CK(8-16), however, changes can seen in their  $\bar{V}_{p3p3}^{T = 0}$ and $\bar{V}_{p3p1}^{T = 0}$ monopole matrix elements due to the above discussed additional modification.

\section{Results and Discussion}
\label{sec3}
\subsection{Effective single-particle energy}
The effective single-particle energy (ESPE), which is more sensitive to the TBMEs and single-particle energies of the orbitals, a good probe to test the basics aspects of effective interaction. The expression of the effective single-particle energy of orbit j$'$ is given as \cite{smirnova}
\begin{equation}
	\epsilon^{'\rho'}_{j^{'}}(A)= \epsilon^{\rho'}_{j^{'}}+ \sum_{j} \hat{n}_{j}^{\rho}  \bar{V}_{jj'}^{\rho \rho'}(A),
\end{equation}
where $\bar{V}_{jj'}^{\rho \rho'}(A)$ is mass dependent monopole matrix elements, and $\hat{n}^{\rho}_{j}$ is the number of particle in the valence orbital j. The symbol $\rho$($\rho'$) denotes the type of particle.

\begin{figure}[ht!]
	\centering
	\includegraphics[height= 9.cm, width = 14cm, trim={1.cm 1.5cm -1.0cm 1.5cm}]{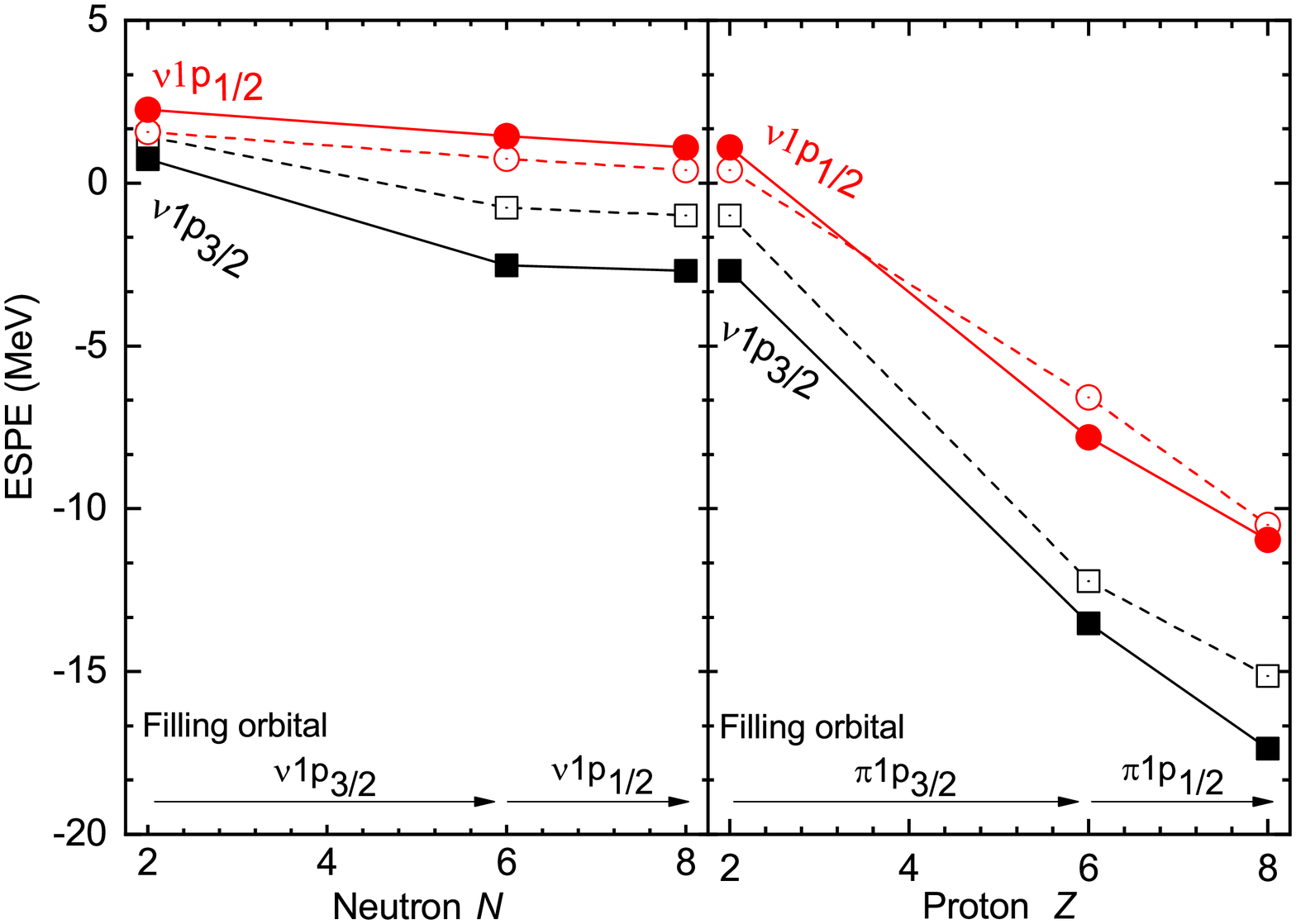}
	\caption{(Color online) Left side: ESPE of $\nu-$1p orbitals with neutron number. Right side: ESPE of $\nu-$1p orbitals in N = 8 isotones from \textit{Z} = 2 to 8. The solid lines and symbols are used for the interaction CKN while dotted lines and open symbols are used for interaction CK(8-16).} 
	\label{F3}
\end{figure}

\begin{table}[h!]
	\centering
	\caption{Contribution of central, spin-orbit and tensor forces of CKN to the orbital energy gap $\nu 1p_{1/2}- \nu 1p_{3/2}$. The unperturbed single-particle energy gap is 1.492 MeV. All numerical values given in the Table are in MeV.}
	\begin{tabular}{lcccccc}
		\hline
		Energy gap    & N = 6 & N =  8 &&& Z =  6 & Z =  8\\
		\cline{2-4}
		\cline{6-7}\\
		Central        &    1.150    &    -0.539  &&&    0.868    &    0.0    \\
		Spin-orbit    &    2.557    &    2.220   &&&    2.106    &    2.613    \\
		Tensor        &    -1.264    &    0.591   &&&    -1.051    &    0.0        \\
		\cline{2-4}
		\cline{6-7}
		Total        &    2.447    &    2.274   &&&    1.922    &    2.613    \\
		\hline
		\\
	\end{tabular}
	\label{tab2}    
\end{table}

\begin{table}[h!]
	\caption{Individual force centroids ($\bar{V}_{jj'}^{T = 1}$) of the interaction CKN. All numerical values given in the Table are in MeV.}
	\centering
	\begin{tabular}{lcccc}
		\hline
		Centroids    &    Central    &    Spin-orbit    &    Tensor    \\
		\hline\\
		$\bar{V}_{p3p3}^{T = 1}$     &    -0.695    &    -0.465    &    0.156    \\
		$\bar{V}_{p3p1}^{T = 1}$    &    -0.341    &    0.322    &    -0.233    \\
		$\bar{V}_{p1p1}^{T = 1}$    &    -1.404    &    0.209    &    0.934    \\
		\\
		Centroids difference ($\Delta$$\bar{V}$) \\
		\hline
		~\\
		$\bar{V}_{p3p3}^{T = 1} - \bar{V}_{p3p1}^{T = 1}$    &    \textbf{-0.354}    &    -0.787    &    \textbf{0.389}    \\
		$\bar{V}_{p3p3}^{T = 1} - \bar{V}_{p1p1}^{T = 1}$    &    \textbf{0.709}    &    -0.674    &    \textbf{-0.778}    \\
		$\bar{V}_{p3p1}^{T = 1} - \bar{V}_{p1p1}^{T = 1}$    &    \textbf{1.063}    &     0.113    &    \textbf{-1.167}    \\
		\hline\\
	\end{tabular}
	\label{tab3}  
\end{table}

Fig.~\ref{F3} shows the variation of ESPEs of $\nu-1p$ orbitals with neutron number(left side) and proton number(right side) from interactions CKN and CK(8-16). The ESPEs of $\nu-1p$ orbitals shows similar trends from both interactions CKN and CK(8-16), and their ESPEs at N = 2 are equal to the unperturbed single-particle energies $\epsilon^{\nu}_{p_{3/2}}$ and $\epsilon^{\nu}_{p_{1/2}}$ as there is no valence neutron present in \textit{p}-model space. 

The $\nu 1p_{1/2}$ and $\nu 1p_{3/2}$ orbitals shown on the left side of Fig.~\ref{F3} goes up when neutrons are removed from the $\nu 1p_{1/2}$ orbital, consequently, the $\nu (1p_{1/2} - 1p_{3/2})$ gap enhances at N = 6. On the right side of Fig.~\ref{F3}, the ESPE of $\nu-$1p orbitals in N = 8 isotones varies from \textit{Z} = 2 to 8 are shown. The gap between the orbitals $\nu 1p_{1/2}$ and $\nu 1p_{3/2} $ is more pronounced at Z = 8, and decreases when the protons are removed from the valence orbitals. As a consequence, the $\nu 1p_{1/2}$ orbital goes up which strongly affects the N = 8 gap between the negative parity orbital $\nu 1p_{1/2}$ and positive parity $\nu 1d_{5/2}$ or $\nu 2s_{1/2}$ orbital, result in the disappearance of N = 8 shell gap and appearance of N = 6 magic shell gap for Z $<$ 6 \cite{sorlin, otsuka7}. This situation is nearly similar to N = 16 magic shell gap in \textit{sd}-shell, which exist for $^{24}_{8}$O and disappear for $^{30}_{14} $Si as $\nu 1d_{3/2}$ orbital goes up when protons are removed from $\pi 1d_{5/2}$ orbital. The stable nucleus $^{8}$He and $^{24}$O with respect to the nucleus $^{9}$He and $^{25}$O are the manifestation of shell gap at N = 6 and 16, respectively. 

In Table.~\ref{tab2}, we summarize the sensitivity of the orbital energy gap $\nu 1p_{1/2}- \nu 1p_{3/2}$ to the individual components of CKN interaction. The spin-orbit force plays a dominant role for $\nu 1p_{1/2}- \nu 1p_{3/2}$ gap in both isotopic and isotonic chain. For $\nu 1p_{1/2}- \nu 1p_{3/2}$ gap at Z = 8, zero contributions from the central and tensor forces shows the important manifestation of the spin-orbit force when both the spin-orbital partners orbitals $\pi 1p_{3/2}$ and $ \pi 1p_{1/2}$ are filled. For $\nu 1p_{1/2}- \nu 1p_{3/2}$ gap at N = 6 and 8, the central and tensor forces are canceling each other contributions due to the almost equal magnitude of their T = 1 centroids with opposite sign, shown in Table.~\ref{tab3}. For $\nu 1p_{1/2}- \nu 1p_{3/2}$ gap at Z = 6, the tensor force is half of the spin-orbit force and lowers the gap. Since the tensor force interaction between $\pi 1p_{3/2}$ and $\nu 1p_{3/2}$ ($j_{>}$ and $j'_{>}$) is repulsive, and between $\pi 1p_{3/2}$ and $\nu 1p_{1/2}$ ($j_{>}$ and $j'_{<}$) is attractive, therefore, the net result in the lowering of the $\nu 1p_{1/2}- \nu 1p_{3/2}$ gap at Z = 6. In Ref. \cite{umeya}, it has been suggested that, the tensor force plays no role if both the spin-orbit partners orbitals are filled. In the present case, the tensor force shows similar behavior for the $\nu 1p_{1/2}- \nu 1p_{3/2}$ gap at Z = 8 when both spin-orbital partners $\pi 1p_{3/2}$ and $\pi 1p_{1/2}$ are filled with protons. 

\begin{figure}[ht!]
	\centering
	\includegraphics[height= 10.0cm, width = 14cm, trim={1.5cm 1.5cm -1.0cm 0.0cm}]{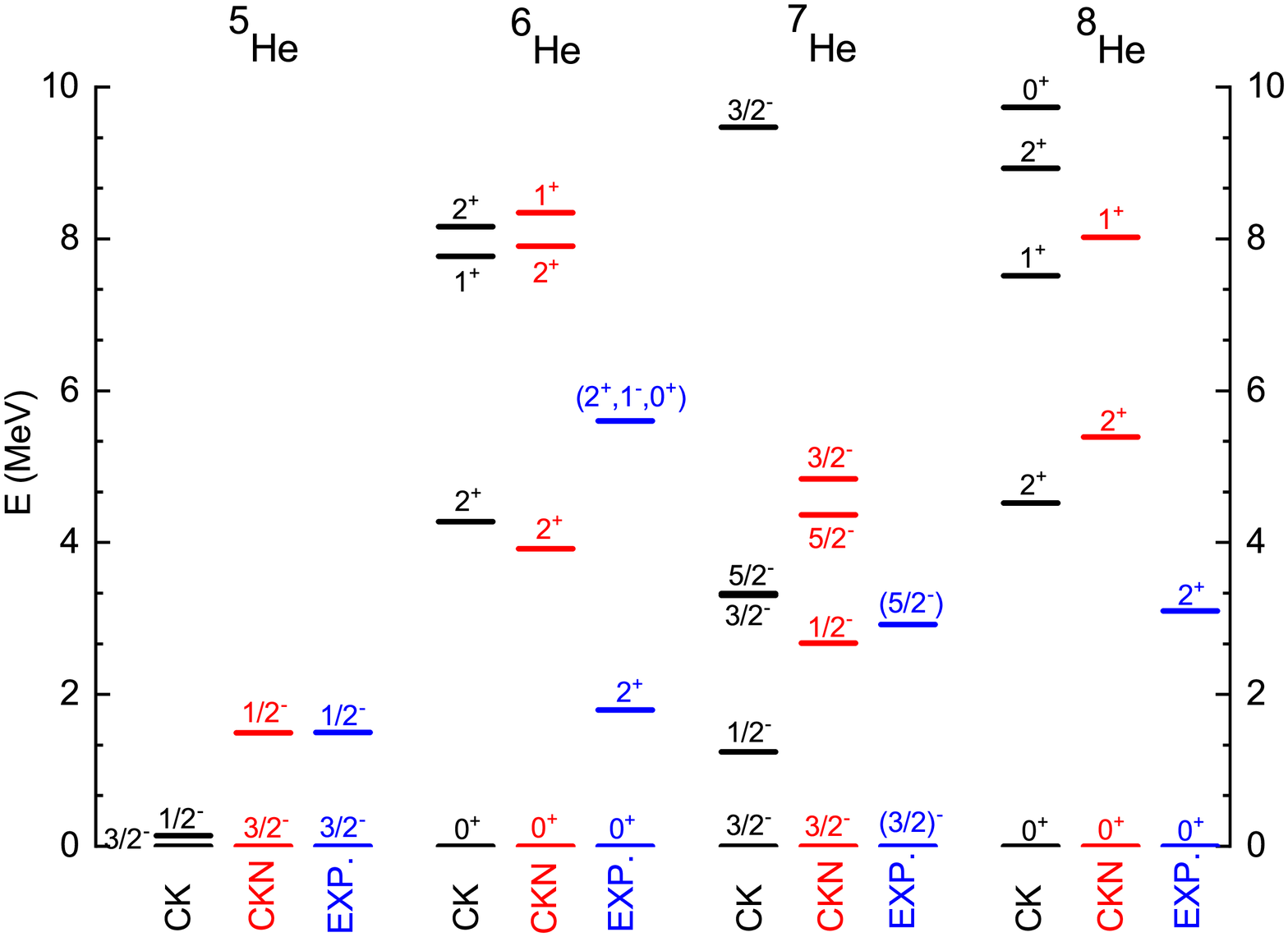}
	\caption{(Color online) Level structure of $^{5-8}$He isotopes. Theoretical calculations are performed with interactions CK(8-16) and CKN. The experimental data are taken from Ref.~\cite{nndc}. The interaction CK(8-16), in short, is represented by CK.} 
	\label{F4}
\end{figure}
\begin{figure}[ht!]
	\centering
	\includegraphics[height= 10.0cm, width = 14cm, trim={1.5cm 1.5cm -1.0cm 0.0cm}]{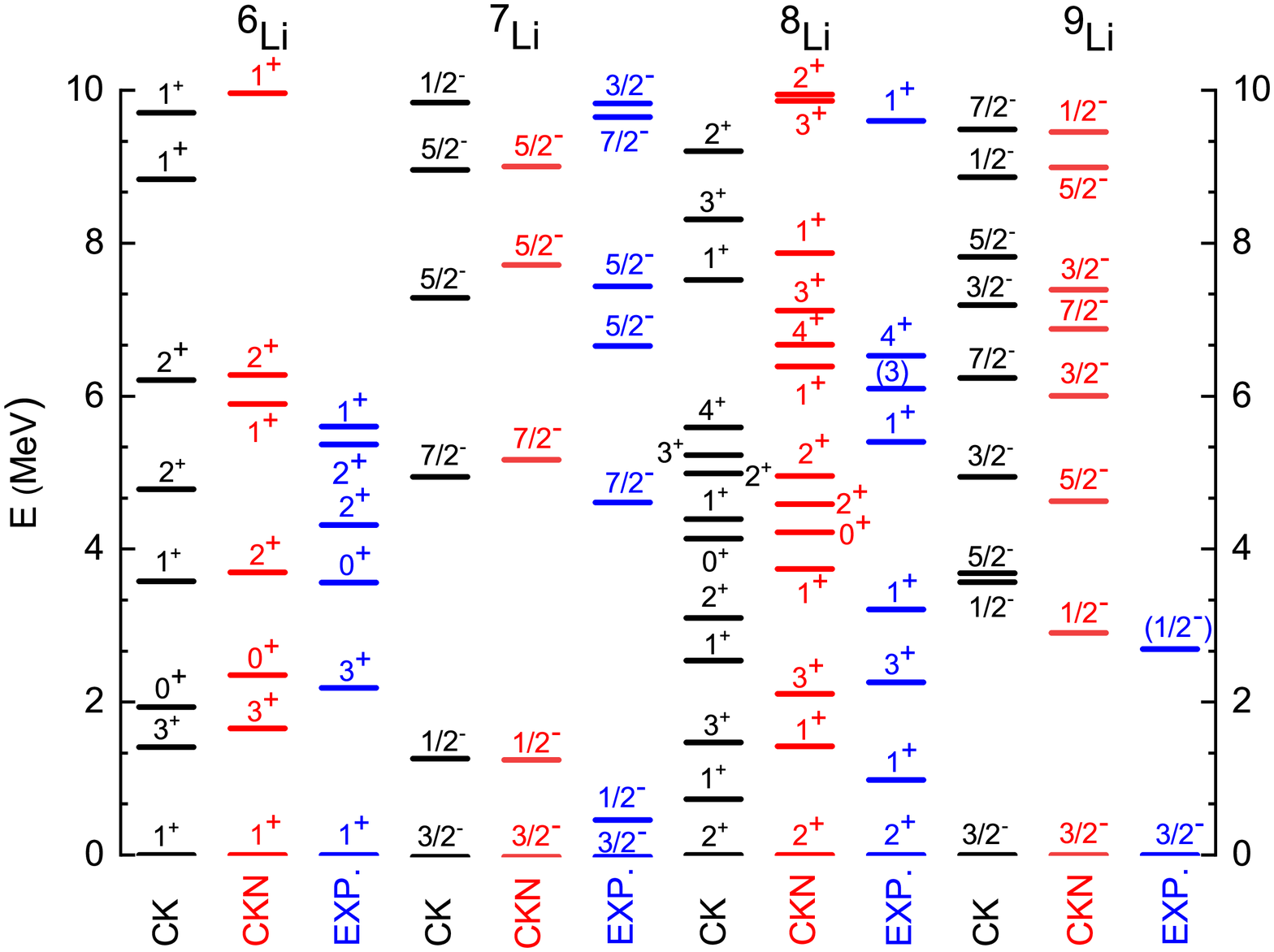}
	\caption{(Color online) Level structure of $^{6-9}$Li isotopes. See caption of Fig.\ref{F4} for details.} 
	\label{F5}
\end{figure}
\begin{figure}[ht!]
	\centering
	\includegraphics[height= 10.0cm, width = 14cm, trim={1.5cm 1.5cm -1.0cm 0.0cm}]{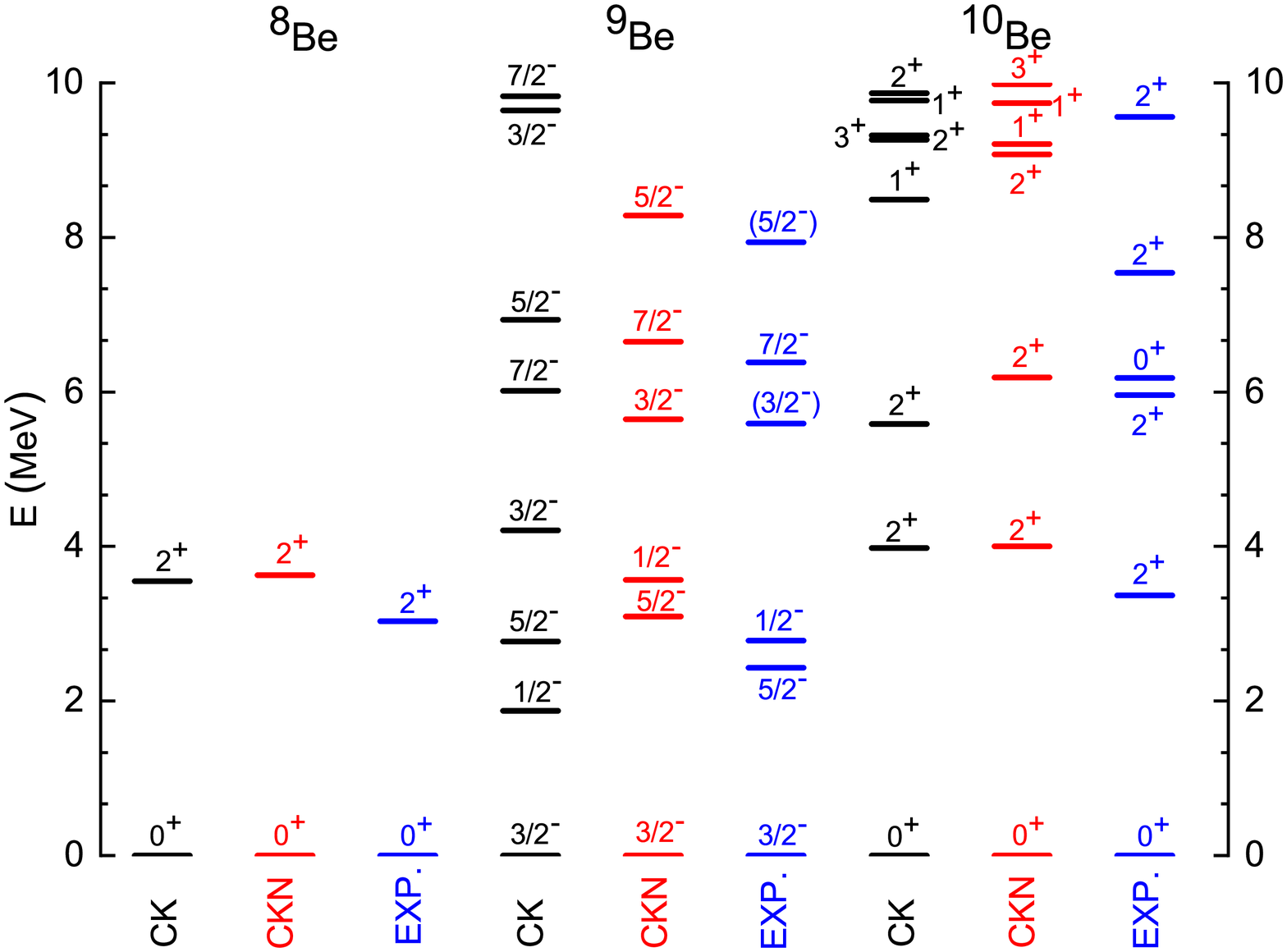}
	\caption{(Color online) Level structure of $^{8-10}$Be isotopes. See caption of Fig.\ref{F4} for details. .} 
	\label{F6}
\end{figure}
\begin{figure}[ht!]
	\centering
	\includegraphics[height= 10.0cm, width = 14cm, trim={1.5cm 1.5cm -1.0cm 0.0cm}]{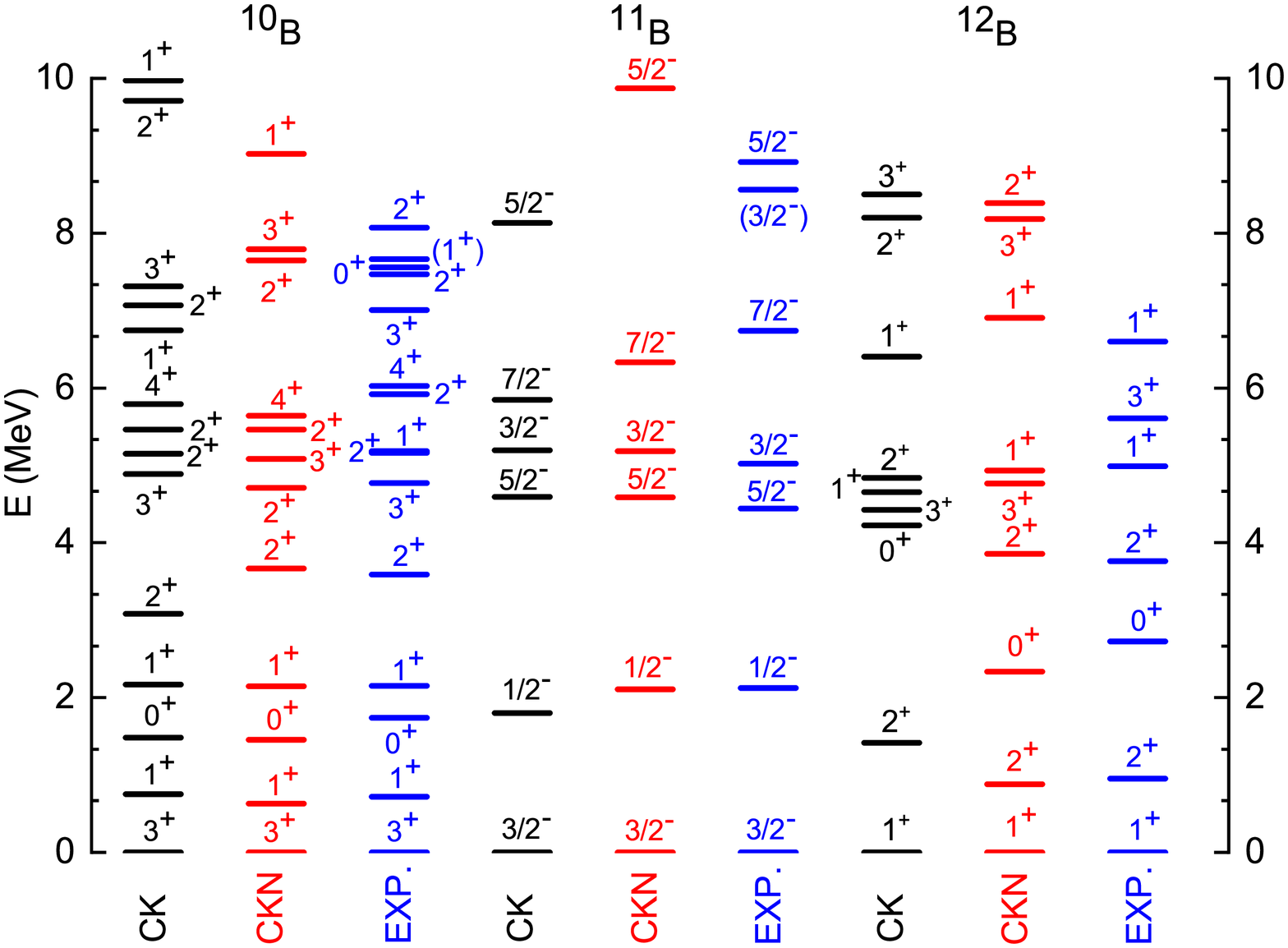}
	\caption{(Color online) Level structure of $^{10-12}$B isotopes. See caption of Fig.\ref{F4} for details.} 
	\label{F7}
\end{figure}
\begin{figure}[ht!]
	\centering
	\includegraphics[height= 10.0cm, width = 14cm, trim={1.5cm 1.5cm -1.0cm 0.0cm}]{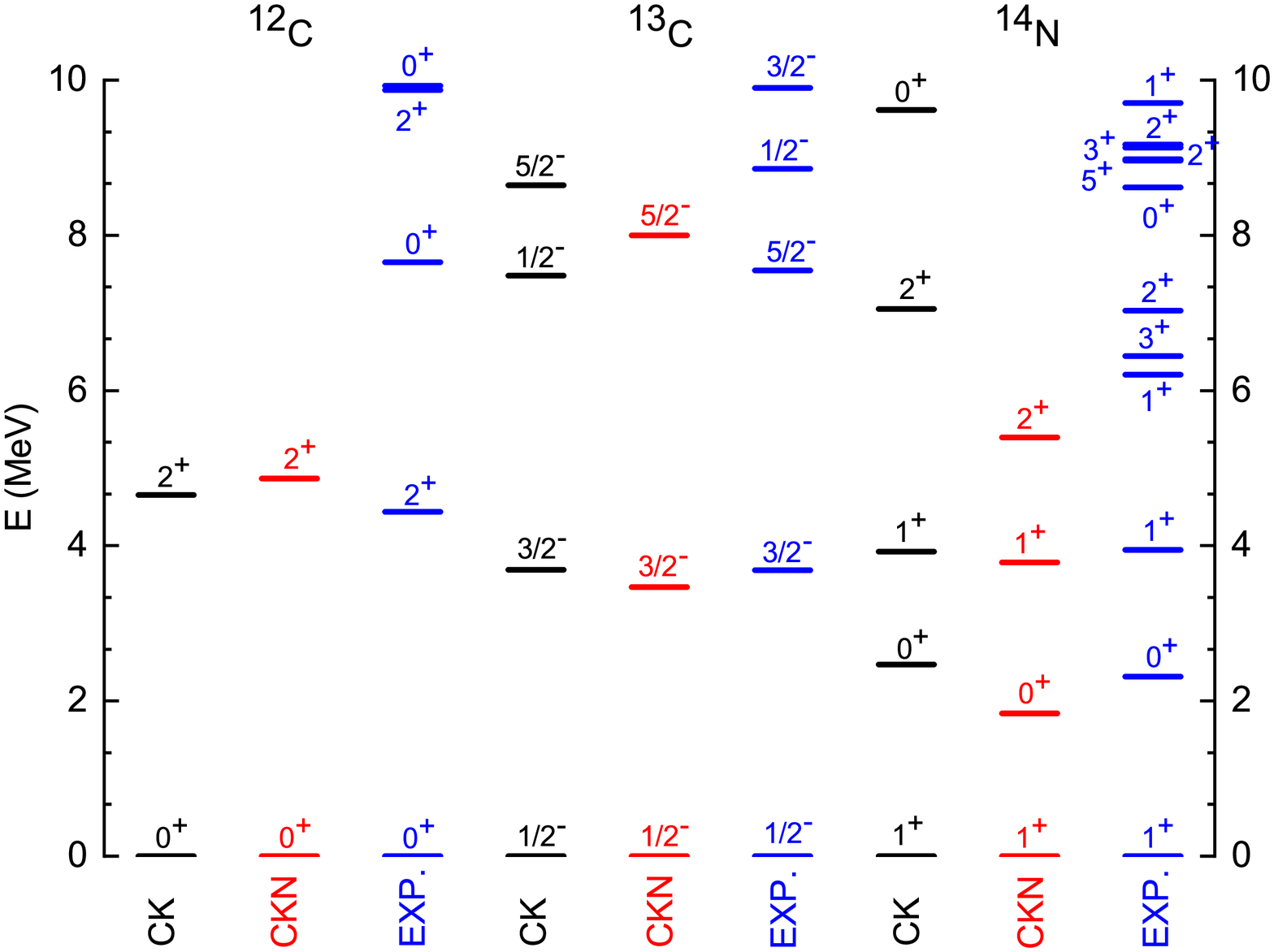}
	\caption{(Color online) Level structure of $^{12-13}$C and $^{14}$N. See caption of Fig.\ref{F4} for details. .} 
	\label{F8}
\end{figure}

\subsection{Level structure calculation}

The excitation energies of the nuclear states are experimentally most accessible to compare the calculated level structure, hence, the level structure data are conventionally used to test the predictive power of the interaction. The level structure of various \textit{p}-shell nuclei of normal parity states are calculated using effective shell-model interactions CKN and CK(8-16), and the obtained results are compared with the experimental data. The theoretical calculations have been performed with shell-model code NUSHELLX@MSU \cite{code}. 

Fig.\ref{F4} shows the level structure for $^{5-8}$He isotopes. For $^{5}$He, the excitation energy of $\frac{1}{2}_{1}^{-}$ state predicted by CKN is in good correspondence with the experimental energy level while it is predicted very low by CK(8-16) interaction. The reason for this is the low single-particle energy gap $\epsilon_{p_{1/2}}-\epsilon_{p_{3/2}}$ of 0.14 MeV in CK(8-16). The 2$_{1}^{+}$ excitation energy of $^{6,8}$He is overpredicted by both the interactions, however, theoretical calculation using CKN reasonably predicts the enhancement of 2$_{1}^{+}$ states from $^{6}$He to $^{8}$He. The other levels of He isotopes are fairly reproduced and close to the calculation done in Ref.\cite{myo}.

Fig.\ref{F5} shows the level structure of $^{6-9}$Li. The 3$_{1}^{+}$ and 0$_{1}^{+}$ states of $^{6}$Li measured at 2.186 MeV and 3.56 MeV are predicted  at 1.658 MeV and 2.355 MeV by CKN, respectively. These states are predicted much below by interaction CK(8-16). The difference between the CKN and CK(8-16) is apparent in 1$_{1}^{+}$ state of $^{6}$Li. This state measured at 5.6 MeV is predicted at 5.899 MeV by CKN while the interaction CK(8-16) predicts it at 3.578 MeV. Likewise in the $^{6}$Li, the excitation energy of 3$_{1}^{+}$ state of $^{8}$Li has been increasing significantly by CKN and it gets closer to the measured value. Further, the interaction CKN remarkably predict 4$_{1}^{+}$ state of $^{8}$Li at 6.67 MeV which is quite close to the experimental value 6.53 MeV. The $\frac{1}{2}_{1}^{-}$ state of $^{9}$Li having configuration  $\pi p_{1/2}^{1} \otimes \nu p_{3/2}^{4}$ with 59.08\% proton single-particle strength. It shows that the $\frac{1}{2}_{1}^{-}$ state of $^{9}$Li are originated from transition of proton from $\pi p_{3/2}$ to $\pi p_{1/2}$ orbital. Further, the $\frac{1}{2}_{1}^{-}$ state of $^{9}$Li is predicted by $\approx $ 877 KeV higher than measured value in CK(8-16) get improved by interaction CKN. 

The level structure of $^{8-10}$Be isotopes are presented in Fig.\ref{F6}. As expected, we find that the calculated ground and excitation energies using CKN are in good correspondence with the experiment. The result obtained from interactions CKN and CK(8-16) are similar for 2$_{1}^{+}$ of $^{8, 10}$Be, although it is predicted higher than the experiment. For $^{9}$Be, the interaction CKN predicts $\frac{3}{2}_{1}^{-}$ and $\frac{5}{2}_{2}^{-}$ states quite closer to the experimental value, and improve the level ordering of  $\frac{5}{2}_{1}^{-}$ and $\frac{1}{2}_{1}^{-}$ states. 

Fig.\ref{F7} shows the level structure of $^{10-12}$B isotopes. The interaction CKN is fairly well reproducing the ground and excites states spectra and maintain the experimentally measured level ordering. The excitation energy of 2$_{1}^{+}$ state of $^{10}$B measured at 3.587 MeV is very nicely predicted at 3.665 MeV by interaction CKN while this state is predicted $\approx $ 500 KeV lower by CK(8-16).

The $\frac{1}{2}_{1}^{-}$ and $\frac{7}{2}_{1}^{-}$ states of $^{11}$B are in good correspondence with the experimental data with interaction CKN than the CK(8-16). These $\frac{1}{2}_{1}^{-}$ and $\frac{7}{2}_{1}^{-}$ states are predicted higher using CKN by 306 KeV and 484 KeV, respectively. For $^{12}$B, the interaction CKN gives better predication for the states $2_{1}^{+}$, $0_{1}^{+}$ and $2_{2}^{+}$ than CK(8-16). These $0_{1}^{+}$ and $2_{2}^{+}$ of $^{12}$B are overpredicted from CK(8-16) by 1.502 MeV and 1.077 MeV, respectively. In Fig.\ref{F8}, we have shown the level structure of $^{12, 13}$C and $^{14}$N. The calculated ground and low excited states of these nuclei are in good agreement with the experiment. The calculated $\frac{5}{2}_{1}^{-}$ state of $^{13}$C by CKN is closed to the measured value.

There are many recent theoretical studies for \textit{p}-shell nuclei using ab-initio approaches \cite{cenxi, navratil, lisetskiy, barrett, jurgenson}. These studies reasonably describe the structural properties of \textit{p}-shell nuclei if three-nucleon force is included in the calculations. In the present work, though, we have adopted a semi-empirical method based on systematic properties to modify an effective interaction, but, it seems to be in the right direction. In the new interaction, the tensor force monopole matrix elements have their systematic features, and from the level structure calculation shown above we can conclude that the overall agreement between the theoretical calculation using CKN with the experiment is quite satisfactory for normal parity states.\\

\subsection{Electromagnetic observables}

The interaction CKN is fairly well describe the excitation spectra of \textit{p}-shell nuclei of normal parity states. In this section, we will investigate the wave functions of it. Since the electromagnetic observables are good probe to test the wave functions \cite{suhonen}, therefore, we have carried out calculations of electromagnetic properties of various \textit{p}-shell nuclei of normal parity states using interactions CKN and CK(8-16), and compared with experimentally available data. The theoretical calculation have been performed using NUSHELLX@MSU code and experimental data are taken from Ref.\cite{nndc, stone}. The measured electromagnetic moments of various \textit{p}-shell nuclei along with the calculated ones using interactions CKN and CK(8-16) are shown in Table.~\ref{tab4}. The magnetic dipole moment operator used for the calculation is 
\begin{equation}
	\mu_{z} (i) = \frac{\mu_{N}}{\hbar} (g_{s}(i) s_{z,i} + g_{l}(i) l_{z,i}),
\end{equation} 
where $l$ and $s$ denote the orbital and spin angular momenta and their corresponding g factors are $ g_{s}$ and $ g_{l}$, respectively. The magnetic dipole moment is defined as the expectation value of the dipole operator \cite{gerda}
\begin{equation}
	\mu =  <J; M = J |\sum_{i} \mu_{z}(i)  | J; M = J>
\end{equation}

The magnetic dipole moments are calculated with the bare value of $g_{s}$ = 5.586 and -3.826, and $g_{l}$ = 1 and 0 for protons and neutrons, respectively. The experimental magnetic dipole moments are in reasonably good agreement with the calculated ones with the bare $g_{s}$ and $g_{l}$. The calculated magnetic moments using interaction CKN for $^{8}$Li, $^{9}$Be and $^{11}$B are quite close to the experimental data, while a large difference can be seen between the measured and calculated ones using interaction CK(8-16). The $\mu (3/2^{-})$ of $^{7}$Li and $^{9}$Li are almost similar due to same spin value; however, a sudden jump in the magnetic moment from $^{6}$Li to $^{7}$Li is indicative of structural change. For $^{10}$Be, the experimental value of  $\mu (2^{+})$ is not known whereas in the theoretical calculation it comes at 1.858 $\mu_{N}$ and 1.787 $\mu_{N}$ by CKN and CK(8-16), respectively. Further, the interaction CKN shift the $\mu (1^{+})$ of $^{14}$N in the right direction towards the experimental value. The rms deviation between theory and experiment is $\pm$ 0.080 $\mu_{N}$ by CKN while it is $\pm$ 0.168 $\mu_{N}$ by CK(8-16) for the data shown in Table.~\ref{tab4}.

The electric quadrupole moment operator used for the calculation is \cite{gerda}
\begin{equation}
	Q_{z} = \sum_{i = 1}^{A}  Q_{z}(i) = \sum_{i = 1}^{A}  e_{i}
	(3z^{2}_{i} - r^{2}_{i}),
\end{equation}
where notation $e_{i}$ represent electric charge and $z_{i}$ and $r_{i} $ are the position coordinates of $i^{th} $ nucleon. The expectation value of $Q_{z}$ gives the spectroscopic quadrupole moment
\begin{equation}
	Q_{s} =  <J; M = J |\sum_{i} Q_{z}(i)  | J; M = J>
\end{equation}

We have calculated electric quadrupole moment ($Q$) of various \textit{p}-shell nuclei with bare value of $e_{p}  = 1.5e $ and $e_{n}  = 0.5e $, shown in Table.~\ref{tab4}. The experimental $Q^{exp}$ are in reasonable agreement with the calculated ones with bare $e_{p} $ and $e_{n}$. For $^{6}$Li, the small quadrupole moment is very nicely reproduced by the interaction CKN. For $^{10}$Be, the calculated quadrupole moment using CKN is close to the experiment and almost twice of the calculated using CK(8-16). We find that the interaction CKN very well reproduce the experimental data except for the case of $^{14}$N where the experimental value is significantly differed by more than two orders of the magnitude than calculated using CKN. The small deviations in quadrupole moment are also commonly seen in the recent studies \cite{suzuki, cenxi}. Further, the CKN successfully reproduce the correct sign of the quadrupole moment for the nuclei discussed in Table.~\ref{tab4}. 

\begin{table}[ht!]
	\caption{Comparison of experimental electromagnetic moments with theoretically calculated ones using interactions CKN and CK(8-16). The experimental data are taken from Ref.\cite{nndc, stone}. The numerical values given in the Table are in $\mu_{N}$ for magnetic dipole moment and $eb$ for electric quadrupole moment.}
	\centering
	\resizebox{\textwidth}{!}{
		\begin{tabular}{lcccccccc}
			\hline
			Nuclei & State &  $\mu^{exp} $  & $\mu^{CKN} $  & $\mu^{CK(8-16)} $ &  $Q^{exp}$ & $Q^{CKN} $ & $Q^{CK(8-16)} $ \\
			\cline{1-8}\\
			$^{6}$Li    &    1$^{+}$    &     +0.8220473(6)    &    0.877    &    0.824    &    -0.000806(6)    &    -0.0007    &    -0.0142    \\
			$^{7}$Li    &    3/2$^{-}$    &    +3.256427(2)        &    3.22    &    3.235    &    -0.0403(4)        &    -0.0382    &    -0.385    \\
			$^{8}$Li    &    2$^{+}$    &    +1.65356(2)        &    1.503    &    1.377    &    0.0326(5)        &    0.0292    &    0.0264    \\ 
			$^{9}$Li    &    3/2$^{-}$    &    3.43678(6)        &    3.445    &    3.471    &    -0.0304(2)        &    -0.0415    &    -0.0421    \\
			$^{8}$Be    &    2$^{+}$    &        -            &    1.009    &    1.007    &     -            &    -0.0783    &    -0.0787    \\
			$^{9}$Be    &    3/2$^{-}$    &    -1.177432(3)     &    -1.191    &    -1.288    &    0.0529(4)    &    0.0487    &    0.0447    \\
			$^{10}$Be    &    2$^{+}$    &        -                &    1.858    &    1.787    &   -0.08(7)        &    -0.0587    &    -0.0272    \\
			$^{10}$B    &    3$^{+}$    &    +1.80064478(6)    &    1.831    &    1.811    &    0.0845(2)    &    0.0897    &    0.0912    \\
			$^{11}$B    &    3/2$^{-}$    &    2.6886489(10)    &    2.716    &    2.534    &    +0.04059(10)        &    0.0486    &    0.0514    \\
			$^{12}$B    &    1$^{+}$    &    +1.00(2)    &    0.828    &    0.599    &    0.0132(3)        &    0.0236    &    0.0189    \\
			$^{12}$C    &    2$^{+}$    &    -                &    1.017    &    1.017    &    +0.06(3)    &    0.0815    &    0.0822    \\
			$^{13}$C    &    1/2$^{-}$    &    +0.7024118(14)        &    0.769    &    0.700    &    -    &    0.0        &    0.0    \\
			$^{14}$N    &    1$^{+}$    &    +0.40376100(6)        &    0.38    &    0.326    &    0.02044(3)        &    0.0007    &    0.0184    \\
			\hline    
		\end{tabular}
	}
	\label{tab4}
\end{table}

In order to get more theoretical insight into the structure, we have calculated electromagnetic transition probabilities of various \textit{p}-shell nuclei. The normal parity states of initial state (\textit{i}) and final state (\textit{f}) are considered in the calculations. The transition probabilities shown in Table.~\ref{tab5} are in Weisskopf unit (W. u.) \cite{suhonen}.  
The transition probabilities calculated using CKN are reasonably good correspondence with the experiment except for some cases where discrepancy still remains. The CKN overestimate the transition $B(M1; 0^{+} \rightarrow 1^{+} )$ of $^{6}$Li while it is underestimated by CK(8-16). In addition, the $B(E2; 2^{+} \rightarrow 1^{+} )$ of $^{6}$Li calculated by CKN has well reproduced the experimental value, whereas it comes almost half by CK(8-16). A similar observation in case of transition $B(E2; 1^{+} \rightarrow 2^{+} )$ of $^{8}$Be. The measured  $B(M1; 1^{+} \rightarrow 2^{+} )$ of $^{8}$Li at 2.8 (9) W.u. is well predicted at 2.752 W.u by CKN. The difference between calculated and experimental values are more than double for $B(E2; 3^{+} \rightarrow 1^{+} )$ of $^{6}$Li and $B(E2; 7/2^{-} \rightarrow 3/2^{-} )$ of $^{11}$B, shown in Table.~\ref{tab5}. For these transitions, initial and final states have a difference in neutron occupation numbers due to the migration of neutron from $\nu 1p_{1/2} $ orbital to the $\nu 1p_{3/2} $ orbital. For all other cases, the transition probabilities are almost similar from both interactions CKN and CK(8-16).

\begin{table}[ht!]
	\centering
	\caption{Electromagnetic transition probabilities in \textit{p}-shell. The shell-model calculations are performed using interactions CKN and CK(8-16), and experimental data are taken from Ref.\cite{nndc}. The E$_{i}$ and E$_{f}$ are the theoretical excitation energies of initial nuclear state (\textit{i}) to a final nuclear state (\textit{f}), respectively. The numerical values given in the Table are in MeV for excitation energies,  and in W.u. for electromagnetic transition probabilities.}
	\resizebox{\textwidth}{!}{
		\begin{tabular}{lccccccc}
			\hline
			Nucleus & E$_{i}$ & E$_{f}$  & J$^{\pi}_{i}$ $\rightarrow$ J$^{\pi}_{f}$ & Multipole & Exp.  & CKN  & CK(8-16) \\
			\hline\\
			$^{6}$Li    &    2.355    &    0    &    0$^{+}$    $\rightarrow$    1$^{+}$    &    M1    &    8.62    (18)    &    9.210 &    8.160    \\
			&    1.658    &    0    &    3$^{+}$    $\rightarrow$    1$^{+}$    &    E2    &    16.5    (13)    &    7.220 &    7.455    \\
			&    3.695    &    0    &    2$^{+}$    $\rightarrow$    1$^{+}$    &    E2    &    6.8    (35)    &    6.926 &    3.593    \\
			$^{7}$Li    &    1.263    &    0    &    1/2$^{-}$    $\rightarrow$    3/2$^{-}$    &    M1    &    2.75    (14)    &    2.419 &    2.370    \\
			&    5.187    &    0    &    7/2$^{-}$    $\rightarrow$    3/2$^{-}$    &    E2    &    4.3                &    6.809 &    6.760    \\
			$^{8}$Li    &    1.419    &    0    &    1$^{+}$    $\rightarrow$    2$^{+}$    &    M1    &    2.8    (9)    &    2.752 &    0.344    \\
			&    2.106    &    0    &    3$^{+}$    $\rightarrow$    2$^{+}$    &    M1    &    0.29    (13)    &    0.399 &    0.356    \\
			$^{8}$Be    &    14.744    &    3.629    &    1$^{+}$    $\rightarrow$    2$^{+}$    &    M1    &    0.17    (4)    &    0.003 &    0.000    \\
			&    14.744    &    3.629    &    1$^{+}$    $\rightarrow$    2$^{+}$    &    E2    &    0.23    (10)    &    0.220 &    0.119    \\
			$^{9}$Be    &    3.095    &    0    &    5/2$^{-}$    $\rightarrow$    3/2$^{-}$    &    M1    &    0.3    (3)    &    0.253 &    0.211    \\
			&    3.095    &    0    &    5/2$^{-}$    $\rightarrow$    3/2$^{-}$    &    E2    &    24.4    (18)    &    21.814  &    20.074    \\
			&    6.652    &    0    &    7/2$^{-}$    $\rightarrow$    3/2$^{-}$    &    E2    &    8.5    (36)    &    7.138  &     6.759    \\
			$^{10}$Be    &    4.002    &    0    &    2$^{+}$    $\rightarrow$    0$^{+}$    &    E2    &    8.00    (76)    &    10.630   &    9.937    \\
			$^{10}$B    &    1.457    &    0.629    &    0$^{+}$    $\rightarrow$    1$^{+}$    &    M1    &    4.2    (18)    &    8.126 &    6.367    \\
			&    3.664    &    0    &    2$^{+}$    $\rightarrow$    3$^{+}$    &    M1    &    0.00026    (15)    &    0.007 &    0.001    \\
			&    0.629    &    0    &    1$^{+}$    $\rightarrow$    3$^{+}$    &    E2    &    3.24    (16)    &    1.908 &    5.571    \\
			$^{11}$B    &    2.109    &    0    &    1/2$^{-}$    $\rightarrow$    3/2$^{-}$    &    M1    &    0.58    (2)    &    1.116 &    1.046    \\
			&    4.590    &    0    &    5/2$^{-}$    $\rightarrow$    3/2$^{-}$    &    M1    &    0.29    (3)    &    0.253 &    0.292    \\
			&    6.334    &    0    &    7/2$^{-}$    $\rightarrow$    3/2$^{-}$    &    E2    &    1.26    (30)    &    3.011 &    3.160    \\
			$^{12}$C    &    12.492    &    4.867    &    1$^{+}$    $\rightarrow$    2$^{+}$    &    M1    &    0.0045    (8)    &    0.001 &    0.001    \\
			&    4.867    &    0    &    2$^{+}$    $\rightarrow$    0$^{+}$    &    E2    &    4.65    (25)    &    9.324 &    9.330    \\
			$^{13}$C    &    3.465    &    0    &    3/2$^{-}$    $\rightarrow$    1/2$^{-}$    &    M1    &    0.39    (4)    &    0.580 &    0.634    \\
			&    3.465    &    0    &    3/2$^{-}$    $\rightarrow$    1/2$^{-}$    &    E2    &    3.5    (8)    &    6.716 &    6.866    
			\\
			\hline
		\end{tabular}
	}
	\label{tab5}   
\end{table}    
We have also calculated \textit{G}amow-\textit{T}eller (\textit{GT}) transition of some of the \textit{p}-shell nuclei in terms of \textit{B}(GT). The \textit{GT} transition between the parent nucleus to a daughter nucleus is one of the sensitive test of the wave functions. The \textit{B}(GT) is defined as \cite{suzuki}
\begin{equation}
	B(GT) = \frac{1}{2J_{i}+1} |<J_{f}|\sigma \tau_{\pm}|J_{i}>|^{2},
\end{equation}
where $J_{i} (J_{f})$ denotes the initial and final state angular momentum, and $\sigma$($\tau$)  represent the spin(isospin) operator. The convention $\tau_{\pm}$ represent $\tau_{+}$$|\nu> $ = $|\pi> $ and $\tau_{-}$$|\pi> $ = $|\nu> $. 

In Table.~\ref{tab6}, we have shown the \textit{B}(GT) of \textit{p}-shell nuclei calculated using interactions CKN and CK(8-16) along with the experimental data. For present purposes, the free value of vector coupling constant $g_{v}$ = 1 and axial-vector coupling constant $g_{a}$ = 1.26 are good enough to calculate \textit{log ft}. The transition $^{6}$He(0$^{+}$)  $\rightarrow$  $^{6}$Li(1$^{+}$) is categories as superallowed transition based on their experimental \textit{log ft} value of 2.9 \cite{suhonen}. In this transition, a 2$-\nu$ nucleus($^{6}$He) having angular momentum zero goes to a $\pi-\nu$ nucleus ($^{6}$Li) having angular momentum one.  The \textit{log ft} value of this transition is predicted by interaction CKN at 2.815, is in good agreement with the experimental value.  It is found that both interactions CKN and CK(8-16) gives almost similar result for the transitions listed in the Table.~\ref{tab6} except for $^{8}$He(0$^{+}$)  $\rightarrow$  $^{8}$Li(1$^{+}$) where CKN is giving better result than CK(8-16). Further, the large difference between the theory and experiment is found for the transitions $^{9}$Li(3/2$^{-}$)  $\rightarrow$  $^{9}$Be(3/2$^{-}$) and $^{12}$Be(0$^{+}$)  $\rightarrow$  $^{12}$B(1$^{+}$), consistent with the recent study \cite{suzuki}. Apart from some exceptions, the proposed interaction very well reproduce the experimental data.

\begin{table}[ht!]
	\centering
	\caption{\textit{B}(GT) transitions in \textit{p}-shell. The shell-model calculations are performed with interactions CKN and CK(8-16). The  experimental data are taken from Ref.\cite{suzuki, navratil}.}
	\begin{tabular}{lccccc}
		\hline
		&  \textit{B}(GT;  $J_{i}^{\pi}$ T $\rightarrow$ $J_{f}^{\pi} $ T)     &  Exp. & CKN     & CK(8-16) \\
		\cline{2-5}\\
		$^{6}$He    $\rightarrow$    $^{6}$Li    &    \textit{B}(GT;  0$^{+}$ 1 $\rightarrow$ 1$^{+}$ 0)    &    4.728(15)    &  5.953    & 5.392     \\
		$^{7}$Be    $\rightarrow$    $^{7}$Li    &    \textit{B}(GT;  3/2$^{-}$ 1/2 $\rightarrow$ 3/2$^{-}$ 1/2)    &    1.3  & 1.621    & 1.616    \\
		$^{7}$Be    $\rightarrow$    $^{7}$Li    &    \textit{B}(GT;  3/2$^{-}$ 1/2 $\rightarrow$ 1/2$^{-}$ 1/2)    &    1.122 & 1.311     & 1.302    \\
		$^{8}$He    $\rightarrow$    $^{8}$Li    &    \textit{B}(GT;  0$^{+}$ 2 $\rightarrow$ 1$^{+}$ 1)    &    0.264(5) &  0.259    &    0.341    \\
		$^{9}$Li    $\rightarrow$    $^{9}$Be    &    \textit{B}(GT;  3/2$^{-}$ 3/2 $\rightarrow$ 3/2$^{-}$ 1/2)    &    0.0190(11) &  0.045    & 0.079    \\
		$^{10}$B    $\rightarrow$    $^{10}$Be    &    \textit{B}(GT;  3$^{+}$ 0 $\rightarrow$ 2$^{+}$ 1)    &    0.08(3)  &  0.036    & 0.147    \\
		$^{12}$Be    $\rightarrow$    $^{12}$B    &    \textit{B}(GT;  0$^{+}$ 2 $\rightarrow$ 1$^{+}$ 1)    &    0.624(3)  &  1.882    & 1.479    \\
		$^{12}$C    $\rightarrow$    $^{12}$B    &    \textit{B}(GT;  0$^{+}$ 0 $\rightarrow$ 1$^{+}$ 1)    &    0.990(2) &  1.033    & 0.992    \\
		\hline
	\end{tabular}
	\label{tab6}
	
\end{table}    

\subsection{Interactions CK(8-16) vs. CKN}
\begin{figure}[ht!]
	\centering
	\includegraphics[height= 10.0cm, width = 14cm, trim={1.5cm 1.5cm -1.0cm 0.0cm}]{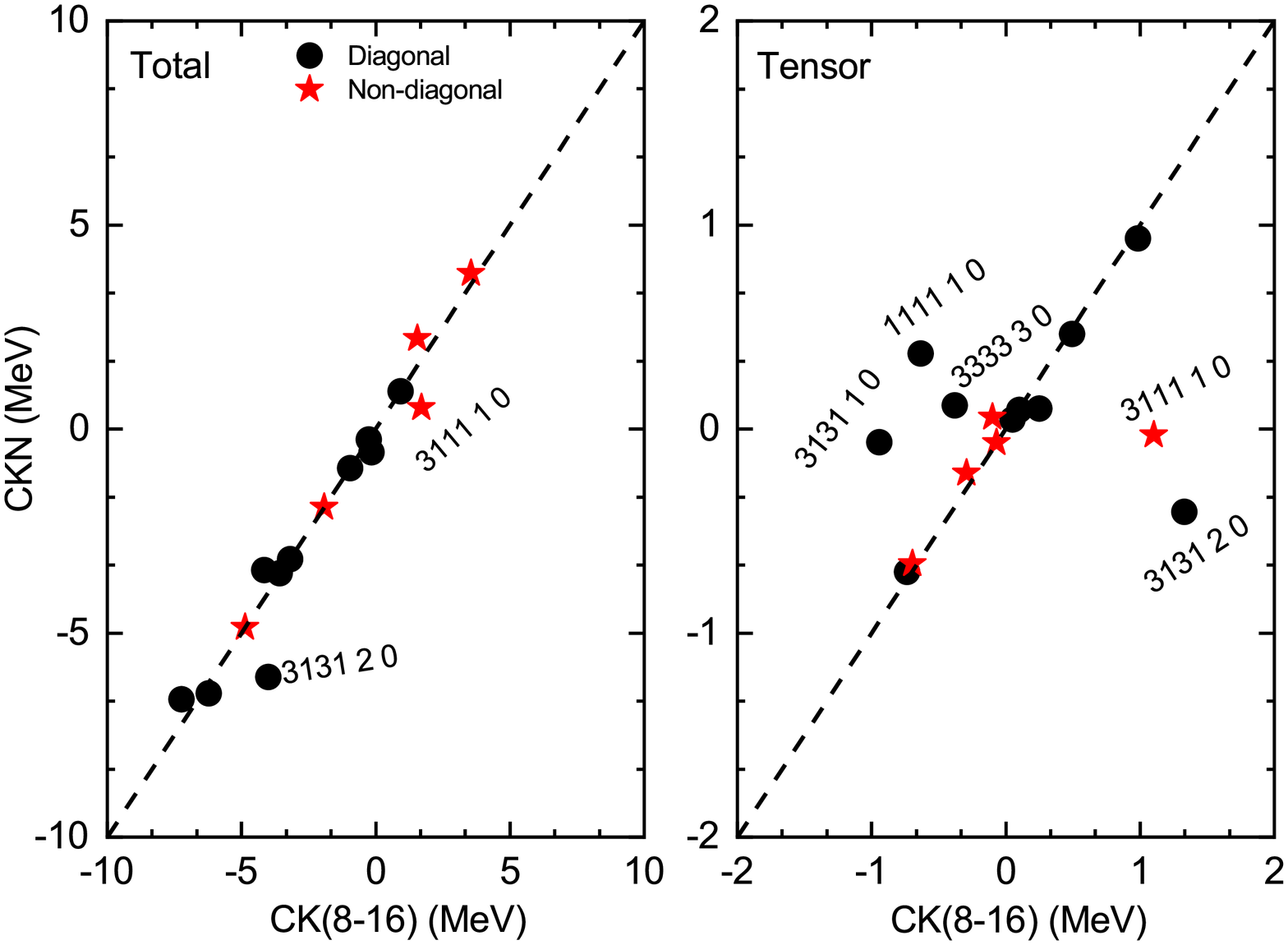}
	\caption{(Color online) Comparison of total and tensor force TBMEs of CKN and CK(8-16). The diagonal and non-diagonal matrix elements are shown by solid circle and solid star, respectively.} 
	\label{F9}
\end{figure}

\begin{figure}[h!]
	\centering
	\includegraphics[height= 8.0cm, width = 12.cm, trim={0cm 1.5cm 0.0cm 0.0cm}]{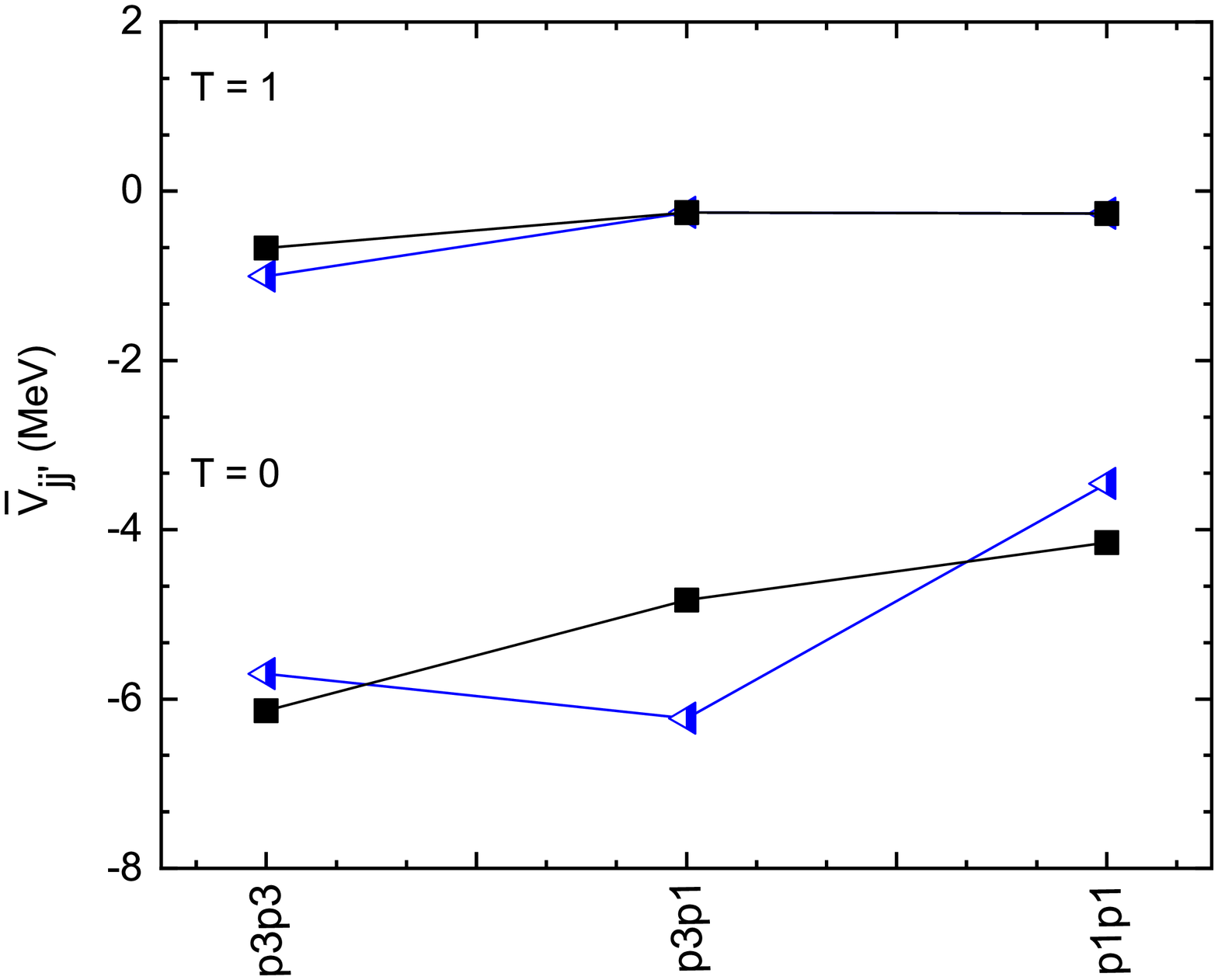}
	\caption{(Color online) Comparison of total monopole matrix elements of interactions CKN (half triangle) and CK(8-16) (solid square) for both Isospin T = 0 and 1. Lines are drawn to guide the eyes.} 
	\label{F10}
\end{figure}

The comparison of total and tensor force matrix elements of interactions CK(8-16) and CKN are shown in Fig.~\ref{F9}. The diagonal and non-diagonal matrix elements of the interactions are shown separately in the figure by the solid circle and solid star, respectively. The tensor force TBME's deviated from the diagonal line are all belong to T = 0, indicates that the T = 1 tensor force TBME's of interactions CK(8-16) and CKN are almost similar. The tensor force matrix elements for which the difference is $\ge$ 0.5 MeV are shown on the right side of Fig.~\ref{F9}. Moreover,  the difference $>$ 1.0 MeV in total matrix elements $ V(3131;20)$ and $ V(3111;10)$ of CKN and CK(8-16) are mainly due to the difference between their corresponding tensor force matrix elements. In Fig.~\ref{F10}, we have shown the comparison of total monopole matrix elements of interactions CKN and CK(8-16) for both Isospin channel T = 0 and 1. The T = 1 monopole matrix elements of both interactions CK(8-16) and CKN are almost similar as expected, however, relatively large difference can be seen for monopole matrix elements $\bar{V}_{p3p1}^{T = 0}$. The $\bar{V}_{p3p1}^{T = 0}$ of CKN is made attractive by -1.39 MeV while $\bar{V}_{p3p3}^{T = 0}$ and $\bar{V}_{p1p1}^{T = 0}$ of CKN are made repulsive by 0.438 MeV and 0.694 MeV, respectively.

Although, the differences in the T = 0  tensor force matrix elements of CK(8-16) and CKN, both the interactions yields almost comparable results in most of the cases discussed. The interaction CKN may be also considered to be a good interaction as far as the normal parity states of \textit{p}-shell are concerned.  

\section{Summary} 
\label{sec4}
In summary, the spin-tensor decomposition method has been used to investigate the tensor force monopole matrix elements properties of interaction CK(8-16), and it is found that their T = 0 matrix elements do not share the systematics trends originating from the bare tensor force. The discrepancies have been corrected using the analytically calculated tensor force matrix elements with some additional modification. The derived interaction obtains this way, named as CKN, is employed to the calculations for \textit{p}-shell nuclei of natural parity states with various physics viewpoints.

To start with, we have analysed single-particle energy variations of $\nu-$1p orbitals with neutron number and proton number. It is found that 
the spin-orbit force plays a dominant role in single-particle energy gap $\nu 1p_{1/2}- \nu 1p_{3/2}$ for both isotopic and isotonic chain, however, the tensor force lowers the $\nu 1p_{1/2}- \nu 1p_{3/2}$ gap at Z = 6 whereas no contribution for gap at Z = 8 in accordance with the nature of  tensor force. For $\nu 1p_{1/2}- \nu 1p_{3/2}$ gap at N = 6 and 8, the net contribution from the central and tensor forces are negligible due to the nearly equal magnitude of their T = 1 centroids with opposite sign. 

The calculated level structure using interaction CKN agree well with the experimental data for \textit{p}-shell nuclei of normal parity states, and we found results are improved for some of the cases. The $3_{1}^{+}$ state of $^{6}$Li, $3_{1}^{+}$ and $4_{1}^{+}$ states of $^{8}$Li,  and $\frac{1}{2}_{1}^{-}$ state of $^{9}$Li have been improved by interaction CKN.  The interaction CKN predicts correct level ordering $\frac{5}{2}_{1}^{-}$ and $\frac{1}{2}_{1}^{-}$ of $^{9}$Be and reproduces correct gap between its $\frac{7}{2}_{1}^{-}$ and $\frac{5}{2}_{2}^{-}$ states. The $2_{1}^{+}$ state of $^{10}$B, $\frac{1}{2}_{1}^{-}$ and $\frac{7}{2}_{1}^{-}$ states of $^{11}$B, and $2_{1}^{+}$, $0_{1}^{+}$ and $2_{2}^{+}$ states of $^{12}$B are in good correspondence with the experimental data with interaction CKN than the CK(8-16). 

The interaction CKN reasonably describe the electromagnetic moments with a few exceptions. The magnetic dipole moments predicted by CKN is quite close to the measured value for $^{8}$Li,  $^{9}$Be,  $^{11}$B, and $^{14}$N and small electric quadrupole moment of $^{6}$Li has been nicely predicted by CKN. We also find good improvement in electromagnetic transition probabilities by CKN specially in case of $B(E2; 2^{+} \rightarrow 1^{+} )$ of $^{6}$Li, $B(M1; 1^{+} \rightarrow 2^{+} )$ of $^{8}$Li and $B(E2; 1^{+} \rightarrow 2^{+} )$ of $^{8}$Be. The \textit{B}(\textit{GT;} $^{8}$He(0$^{+}$)  $\rightarrow$  $^{8}$Li(1$^{+}$)) is also better predicted by CKN than CK(8-16). Despite the major differences in the matrix elements $ V(3131;20)$ and $ V(3111;10)$ of interactions CKN and CK(8-16), the other matrix elements are almost similar,  yields almost comparable results in most of the cases. In the present work, though, we have adopted a semi-empirical method based on systematic properties to modify an effective interaction, but, the modification has improved the predictability of the interaction.

\section*{Acknowledgments}
K. Jha acknowledges Pooja Siwach and R. N. Sahoo for their interest and useful scientific discussion, and he is also grateful to the Ministry of Human Resource and Development, Government of India for providing financial support. 

\end{document}